\documentclass[aps,prb,reprint,superscriptaddress]{revtex4-1}

\usepackage{graphicx, color}
\usepackage{xcolor}
\usepackage{amsfonts,amssymb,amsmath}
\usepackage{siunitx}
\usepackage{bm}
\usepackage{bbold}
\usepackage{braket}
\usepackage{mathtools}
\usepackage{dsfont}
\usepackage{booktabs}

\usepackage[symbol]{footmisc}

\begin{document}

\title{Opportunities in topological insulator devices}

\author{Oliver Breunig}
\affiliation{Physics Institute II, University of Cologne, Z\"ulpicher Str. 77, 50937 K\"oln, Germany}

\author{Yoichi Ando}
\email[]{ando@ph2.uni-koeln.de}
\affiliation{Physics Institute II, University of Cologne, Z\"ulpicher Str. 77, 50937 K\"oln, Germany}

\begin{abstract}
Topological insulators (TIs) are expected to be a promising platform for novel quantum phenomena, whose experimental realizations require sophisticated devices. In this Technical Review, we discuss four topics of particular interest for TI devices: topological superconductivity, quantum anomalous Hall insulator as a platform for exotic phenomena, spintronic functionalities, and topological mesoscopic physics. We also discuss the present status and technical challenges in TI device fabrications to address new physics.
\end{abstract}

\maketitle

\section{Introduction}
After more than 10 years of research, the understanding of topological insulator (TI) materials\cite{Ando2013} has been well advanced. The next step is to use them as a platform for devices to realize novel and useful topological phenomena, such as emergence of chiral Majorana fermions\cite{He2017, Kayyalha2020}, topological qubits using Majorana zero-modes\cite{Aguado2020, Manousakis2017}, or topological magnetoelectric effects\cite{Qi2008} in the axion insulator state\cite{Mogi2017,Xiao2018} (these concepts are explained later). Also, mesoscopic physics of the topological states of matter is a rich realm\cite{Muenning2021}, but it has been largely left unexplored. Hence, TI devices provide promising opportunities for new discoveries.

TIs are characterized by a nontrivial $Z_2$ topology of their bulk electronic wave functions, which leads to the appearance of topologically-protected Dirac surface states\cite{Ando2013}. The most important property of the topological surface states is the spin-momentum locking, which lifts the spin degeneracy and dictates the spin orientation depending on the momentum $\mathbf{k}$. In fact, this property makes TIs a promising platform for topological superconductivity or spintronic devices.

To characterize TIs, the experimental methods that have been employed are transport measurements, angle-resolved photoemission spectroscopy (ARPES), scanning tunneling microscopy / spectroscopy (STM/STS), magneto-optical spectroscopy, and ultrafast optics measurements, among others. The spectroscopic experiments are useful for understanding the energy-, momentum-, spatial-, time-, and spin-dependence of the topological electronic structures. On the other hand, transport measurements are suitable for probing the low-energy physics occurring at the Fermi level, such as quantum oscillations, superconductivity, or quantum Hall effects, and employing device fabrications in transport experiments allows us to have a tuning knob for systematic measurements or to manipulate the electronic states at will.
  
Even though it is clear that TI devices would be useful for the exploration of novel topological phenomena, it is important to note that almost all known TI materials are much less robust against device fabrications compared to typical semiconductor materials such as Si or GaAs. Also, the functionally active part of TIs is the surface states and the bulk conduction should be suppressed as much as possible. These circumstances make it important to apply techniques specifically tuned for TI devices. Such techniques have been reasonably well developed and various kinds of TI devices are being fabricated, allowing for observations of interesting phenomena peculiar to TIs\cite{He2017, Mogi2017, Xiao2018, Muenning2021, Cho2015, Wiedenmann2016, Deacon2017, Taskin2017, Yasuda2017, Chen2018, Schueffelgen2019}. 

In this Technical Review, we discuss important themes to be addressed by using TI devices. They include realization of topological superconductivity, novel topological phenomena based on quantum anomalous Hall insulators, spintronic functionalities of TIs, and topological mesoscopic physics. After these physics discussions, technical challenges in TI devices are discussed and suggestions are made for the realization of these expectations.

 \begin{figure*}[t]
 \centering
 \includegraphics[width=\linewidth]{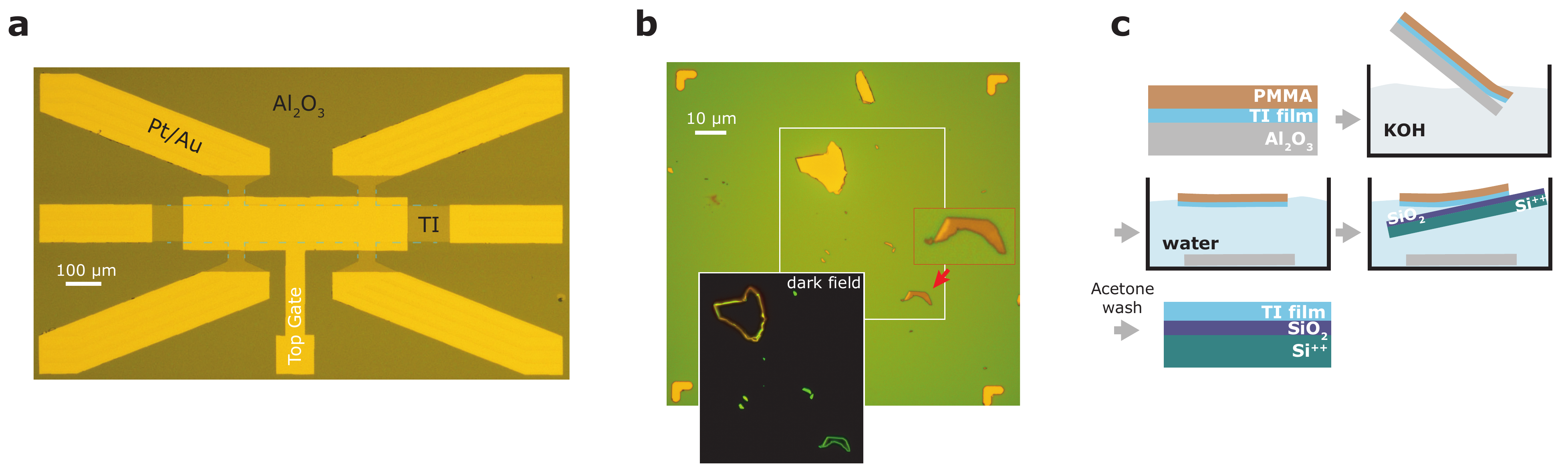}
 \caption{\textbf{Topological insulator device fabrication.} \ 
  \textbf{a}, Top-gated Hall-bar device fabricated from a MBE-grown thin film. The dashed line indicates the edge of the etched Hall bar structure underneath the gate metal.
 \textbf{b}, Optical contrast of thin exfoliated flakes. The flake marked by a red arrow and shown in a magnified view (box framed red) is less than 20\,nm thick and thus appears half-transparent. On its left edge the thickness is larger, indicated by a brighter color. In dark-field microscopy (inset) the edge of thin flakes is less apparent than that of thicker flakes. A dense array of prefabricated markers (corners of the image) is useful for precise alignment.
 \textbf{c}, Process of transferring a TI film from an Al$_2$O$_3$ growth substrate to a Si/SiO$_2$ substrate for dual-gating, following Ref.~\citenum{Yang2015}. 
 }
 \label{fab}
 \end{figure*}

\section{Selected topics of interest in the context of TI Devices}

Besides the basic devices used for characterizations (see Fig.~\ref{fab}a as an example), TI devices are made to realize the conditions required for novel quantum phenomena and/or to address the peculiar properties expected for TIs. The devices to combine TIs with a superconductor or a ferromagnet (or both) are of particular interest because of their prospect for novel applications in quantum computing or spintronics. In this section, we discuss the physics behind these expectations.

\begin{flushleft}
{\bf Topological superconductivity:}
\end{flushleft}
\vspace{-2mm}

Topological superconductivity is a broad concept for superconductors characterized by some non-trivial topological invariant\cite{Sato2017}. While a large part of the recent attention on topological superconductivity is due to its relevance to topological quantum computing, it is useful to note that not all topological superconductors are useful for this purpose. For example, a spinful chiral $p$-wave superconductor (for which Sr$_2$RuO$_4$ is a candidate\cite{Mackenzie2017}) is not suitable. The key ingredient for topological quantum computing is an isolated Majorana zero mode (MZM): When isolated, this mode is pinned to zero energy and can be viewed as a half of an electron, because it requires two MZMs to accommodate an unpaired electron in the topological superconductor\cite{Alicea2010}. This situation in turn means that the information carried by an electron can be stored non-locally in two MZMs that are spatially far apart, which gives an advantage in protecting the quantum information when MZMs are used for encoding a qubit\cite{Aasen2016}. Another, more fundamentally important, advantage of a qubit based on MZMs is the possibility to perform topologically-protected gate operation, which is called ``braiding''\cite{Alicea2010}. Typically, a Majorana qubit consists of four MZMs, and a braiding operation exchanges two of the four MZMs\cite{Aasen2016}. One such braid changes the qubit state $|0 \rangle$ into an equal superposition of  $|0 \rangle$ and  $|1 \rangle$ states, which is equal to a $\pi/2$ rotation of the qubit state on the Bloch sphere. Moreover, two braids brings the $|0 \rangle$ state into the $|1 \rangle$ states, which is a $\pi$ rotation of the qubit state. Theoretically, the result of these one-bit gate operations performed via braiding depends only on the number of braiding and has no timing-error, which is the reason why it is called topological quantum computing\cite{Nayak2008}.

Importantly, in order for the electron condensate to become a topological superconductor capable of harbouring MZMs, it needs to be spin non-degenerate, the property often called ``spinless''. Such a spinless situation is realized when electrons are spin-polarized and only one spin subspace is relevant to the low-energy physics. The ferromagnetic state in a half-metal naturally satisfies such a condition, but also the spin-momentum-locked surface states of TIs are spinless. The advantage of the TI surface states is that the spin is antiparallel between electrons with momentum $\mathbf{k}$ and $-\mathbf{k}$, which makes it easy to induce Cooper pairing between these surface electrons by using the superconducting proximity effect from a conventional $s$-wave superconductor. Still, the spinless nature makes it possible to map the superconducting state induced in the TI surface to the spinless chiral $p$-wave state using a simple unitary transformation\cite{Fu2008}. The latter is a prototypical topological superconducting state which harbours a MZM in the vortex core. Note that the superconducting TI surface is time-reversal-invariant and one needs to break the time-reversal symmetry (e.g. by threading a vortex) to generate MZMs.

\begin{figure*}[t]
 \centering
 \includegraphics[width= 0.75\linewidth]{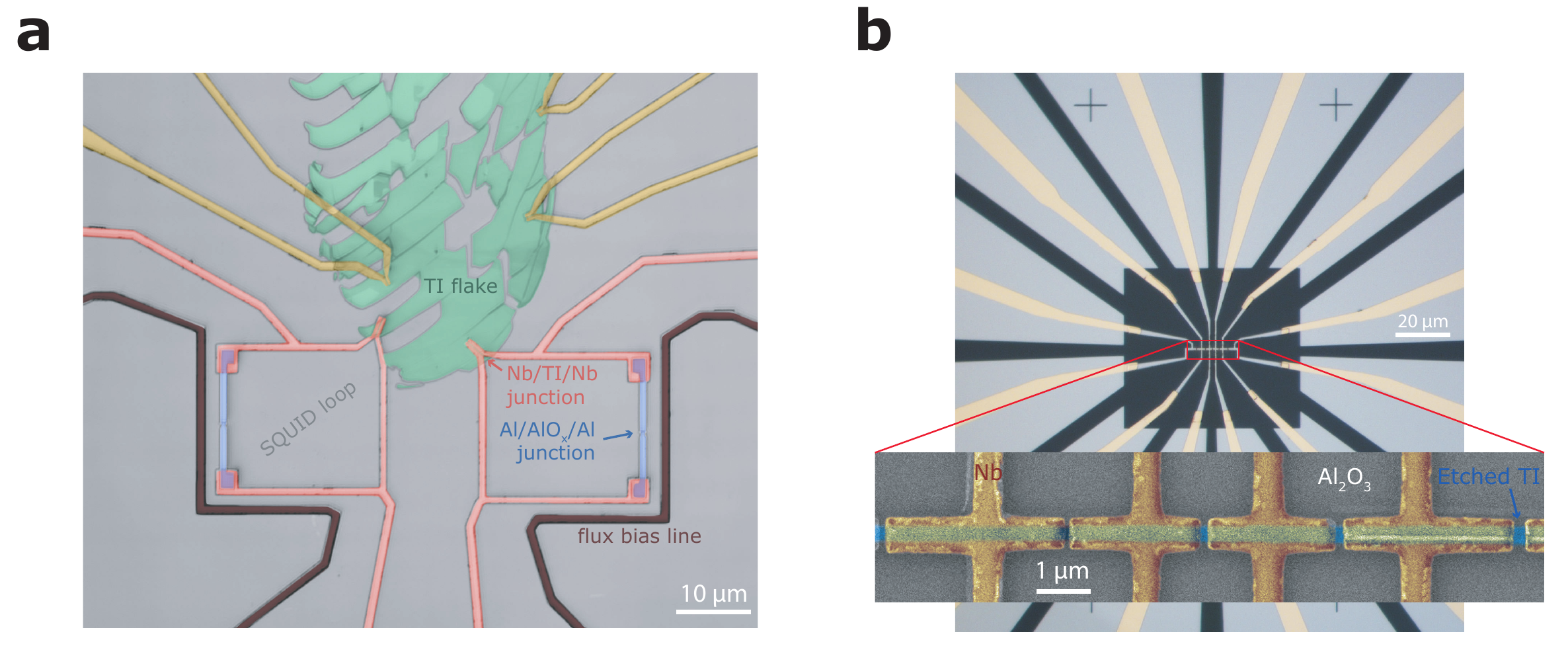}
 \caption{\textbf{Topological insulator/superconductor devices.} \ 
 \textbf{a}, False-color confocal laser micrograph of an asymmetric SQUID-device for investigating the current-phase-relation of a SC/TI/SC Josephson junction. The SQUID loop is formed by an Al/AlO$_x$/Al reference junction (blue) in parallel to a TI junction made from two close Nb leads (light red). A flux bias line (brown) is used to alter the flux through the loop. Auxiliary test junctions (yellow) are fabricated on top of the exfoliated flakes (green) in order to characterize the stand-alone SC/TI/SC junction. 
 \textbf{b}, TI nanowire Josephson junction device fabricated by etching a MBE-grown thin film into a nanowire shape (optical image). The inset shows a close-up false-color scanning electron micrograph of the nanowire (blue shade) with differently spaced Nb leads (gold) on top, forming a series of Josephson junctions that are individually accessible. }
 \label{JJ}
 \end{figure*}

Since MZMs in vortex cores are difficult to access and manipulate in devices, experiments have been performed on TI-based Josephson junctions\cite{Williams2012, Oostinga2013, Ghatak2018, Wiedenmann2016, Bocquillon2017, Li2018, Schueffelgen2019, Deacon2017} (see Fig.~\ref{JJ} as examples). Unless one makes a tri-junction\cite{Fu2008}, MZMs cannot be generated in a Josephson junction. Nevertheless, TI-based Josephson junctions can still harbour dispersive Majorana bound states (MBSs), which is distinguished from the usual Andreev bound states in the $4\pi$-periodic dependence on the phase difference between the two superconducting electrodes\cite{Fu2008}. The $4\pi$-periodicity manifests itself most prominently in the ac Josephson effects, either in the form of missing 1st Shapiro step upon microwave irradiation\cite{Wiedenmann2016, Bocquillon2017, Li2018} or halving of the emitted microwave frequency upon dc voltage application\cite{Deacon2017}. One should keep in mind, however, that an apparent $4\pi$-periodicity can be observed due to excited quasiparticles or Landau-Zener tunneling\cite{Dartiailh2021}. Hence, the Josephson junction experiments can only provide indirect support to the realization of topological superconductivity.

After the ground-breaking proposal by Fu and Kane\cite{Fu2008}, it was recognized that a spinless band structure similar to that in a TI surface can be engineered in a spin-orbit-coupled semiconductor (such as InAs) by combining a Rashba-type band splitting and a Zeeman energy gap\cite{Alicea2010}. By engineering such spinless bands in a one-dimensional (1D) semiconductor nanowire and by employing the superconducting proximity effect, MZMs can be generated at the ends of the nanowire, even though it requires a large magnetic field. This 1D semiconductor platform has been considered to be the main contender for the topological quantum computing and a lot of high-profile experiments have been reported\cite{Aguado2020, Prada2020}. However, unambiguous evidence for MZMs has not been obtained and it was recently pointed out that the density of impurities should be around 10$^{15}$ cm$^{-3}$ or lower to realize a Majorana qubit\cite{Woods2021}; this is about 100 times lower than the current state of the art.

In this regard, it was recently proposed that taking the unavoidable disorder into account, TI nanowires could be a more promising platform for generating robust MZMs; in that proposal\cite{Legg2021}, a lifting of the degeneracy by a combination of an electric field (produced by a gate) and a moderate magnetic field provides the spinless band in a reasonably wide energy window of the order of 10 meV. Here, the quantum confinement of the TI surface states to create a peculiar subband structure (described later) plays a key role, besides the fact that a 1D topological superconductor naturally harbours MZMs at the ends without the need to create vortices.
Nevertheless, the bulk-insulating TI nanowires that are currently available are compensation-doped\cite{Muenning2021}, which leaves charged impurities causing fluctuations of the Coulomb potential. A recent theory predicted\cite{Huang2021} that the impurity density should be around 10$^{18}$ cm$^{-3}$ or lower to realize a Majorana device with TI nanowires, in which the potential fluctuations should remain smaller than the energy window for the spinless band. (Note that the analytic estimate in Ref. \citenum{Huang2021} shows that for the impurity density $\gtrsim 10^{18}$ cm$^{-3}$, the subband features are completely smeared in the transport properties of a nanowire. However, gate-voltage-dependent resistance oscillations reflecting the subband gaps have been experimentally observed\cite{Muenning2021}, suggesting that the impurity density in TI nanowires may actually be lower than this threshold.)

\begin{flushleft}
{\bf Quantum anomalous Hall insulator:}
\end{flushleft}
\vspace{-2mm}

The quantum anomalous Hall (QAH) effect is similar to the quantum Hall effect but can be achieved without the formation of Landau levels in a large magnetic field; instead, it signifies the quantized Berry curvature of the occupied states. This quantization happens when the Fermi level is tuned into the magnetic exchange gap opened at the charge neutrality point (Dirac point) of the TI surface states in the presence of a ferromagnetic order. The QAH effect can be realized in thin films (thickness $\lesssim$ 10 nm) of the bulk-insulating TI material (Bi$_{1-x}$Sb$_x$)$_2$Te$_3$ doped with Cr or V to induce ferromagnetism\cite{Chang2013,Chang2015}. It can also be achieved in thin exfoliated flakes of MnBi$_2$Te$_4$\cite{Deng2020}. In the QAH effect, the Hall conductance $\sigma_{yx}$ is quantized to $e^2/h$ and the longitudinal conductance $\sigma_{xx}$ vanishes. The ferromagnetic TI that shows the QAH effect is called a QAH insulator, because the ``bulk'' of the two-dimensional (2D) surface states are gapped and only the 1D chiral edge states carry the electric current. The physics of these chiral edge states is itself an interesting subject to study. For example, this chiral edge states show up at the magnetic domain boundaries in a QAH insulator and they can be moved by manipulating the magnetic domains by, for example, using the tip of a magnetic force microscope\cite{Yasuda2017}. Also, it was reported that the $\sigma_{xx} = 0$ state is extremely unstable against the electric current\cite{Kawamura2017, Lippertz2021}, which is an indication of an intricate interaction between the edge states and the bulk states. Furthermore, this interaction leads to a large non-reciprocal transport through the edge states\cite{Yasuda2020}.

In a QAH insulator, the magnetization should be parallel in the top and bottom surfaces. By making the magnetization of the two surfaces to be antiparallel, one can realize an axion insulator\cite{Mogi2017,Xiao2018}, in which the whole system (bulk and edge) is gapped. According to theory\cite{Qi2008}, such an axion insulator presents interesting topological magnetoelectric effects: For example, a charge polarization will induce bulk magnetization with the proportionality coefficient given by the fine structure constant $\alpha$ ($= e^2/\hbar c$). 

When a QAH insulator is in contact with a conventional $s$-wave superconductor, the combination of electron doping from the superconductor to the QAH-insulator surface and the superconducting proximity effect is expected to lead to a time-reversal-breaking topological superconductivity\cite{Qi2010}. The existence of ferromagnetism in the QAH-insulator surface makes the situation different from the case of the proximitization of a pristine TI. Due to the time-reversal-breaking, the proximitized QAH insulator is expected to be accompanied by chiral Majorana edge states\cite{Qi2010}, which is a dispersive 1D edge mode having the Majorana character --- namely, this mode is self-conjugate. In quantum field theory, a self-conjugate particle is its own antiparticle, and it is the decisive character of the original Majorana fermions conceived by Ettore Majorana as a model for neutrinos\cite{Wilczek2009}. In this sense, one may say that this 1D edge mode realizes chiral Majorana fermions. Therefore, proximity-induced superconductivity in a QAH insulator is of particular interest in Majorana physics, and there are already experimental works along this line\cite{He2017,Kayyalha2020}. However, the generation of chiral Majorana fermions remains controversial\cite{Kayyalha2020}.

The chiral Majorana fermions are dispersive fermionic quasiparticles that cannot themselves be used for topological quantum computing\cite{Beenakker2019}. Nevertheless, a phase boundary in the chiral Majorana edge state is a zero-mode and can be used for braiding\cite{Beenakker2019}. Such a phase boundary is created upon crossing of a vortex through the edge state. Interestingly, this zero-mode has charge $e/2$ and is carried by the 1D edge mode\cite{Hassler2020}. A Majorana qubit which is based on this type of MZMs is conceivable and it will belong to the ``flying qubit'' category\cite{Beenakker2019}, which has an obvious advantage in transporting quantum information. The braiding experiment in this case requires a rather complex device setup and time-resolved experiments\cite{Beenakker2019,Hassler2020}. While very challenging, such an experiment would be extremely interesting. A different approach to use gating to define 1D channels in a QAH insulator proximity-coupled to a superconductor has also been proposed for generating MZMs\cite{Zeng2018}.

\begin{flushleft}
{\bf Spintronic functionalities:}
\end{flushleft}
\vspace{-2mm}

The spintronic functionalities of TIs arise naturally from the spin-momentum locking of the surface states: One can induce spin polarization by creating electric current in the surface, whereby the orientation of the spin polarization is controlled by the direction of the current. Note that the induced spin polarization in the TI surface is primarily confined within the surface plane, but the trigonal warping term in the surface Hamiltonian gives rise to an out-of-plane spin component\cite{Fu2009Warping}, which can be useful for using TI as a controllable spin source.
 
In the diffusive transport regime, the mechanism of current-induced spin polarization is the Edelstein effect\cite{Hellman2017} --- the electric current creates an imbalance in the electron population on a spin-polarized Fermi surface, resulting in the dominance of one spin orientation on average. The extent of the imbalance depends inversely on the electron scattering rate. Hence, cleaner surface creates larger spin polarization.

The current-induced spin-polarization on the TI surface can be detected with a ferromagnetic electrode. The principle of spin detection is based on the difference in the electrochemical potential for up and down spins created by nonequilibrium spin polarization in the TI surface\cite{Hellman2017,Yang2016}. When a ferromagnet is in contact with the TI surface, the electrochemical potential of the ferromagnet equilibrate with that of the same spin orientation in the TI surface; as a result, the voltage of the ferromagnet with respect to a reference nonmagnetic electrode changes depending on the induced spin polarization on the TI surface. A pictorial explanation of this principle can be found in Ref. \citenum{Yang2016}. Note that, due to the negative charge of electrons, the magnetization on the TI surface is antiparallel to the current-induced spin polarization. There has been a confusion\cite{Li2014} in data interpretation caused by this sign difference.

It is useful to mention that even though the helical spin polarization of the Dirac surface states changes sign across the Dirac point, the current-induced spin polarization is {\it not} expected to change sign. This is because for a given $\mathbf{k}$, not only the spin but also the Fermi velocity change sign upon crossing the Dirac point. These two sign changes cancel each other in the Edelstein effect and lead to the same spin polarization. Nevertheless, if Rashba-type surface states are created by a surface band bending and coexist with the topological surface states, their competition might lead to a sign change in the current-induced spin polarization upon gating, leading to a ``spin transistor'' operation\cite{Yang2016}.

Even though the controllability of the spin polarization via current is appealing, the efficiency of spin generation on the TI surface is inherently low in the diffusive transport regime, as is always the case with the Edelstein effect for spin generation. In this regard, there is an interesting possibility for TIs to dramatically increase the spin generation efficiency: In the ballistic transport regime, the spin-momentum locking means that a 100\% spin polarization is expected in theory. Hence, TI-based nanodevices to pursue this avenue would be very useful. 
It should me mentioned, however, that TI nanowires may not be suitable for this purpose, because the 1D bands formed in TI nanowires are spin-degenerate, as explained in the next section. A magnetic flux can lift the spin degeneracy, but it may cause a problem in spin detection. Also, the mean free path of currently-available TI nanowires is $\sim$100 nm and true ballistic transport has not been achieved\cite{Cho2015}; as long as the transport is diffusive, the Edelstein effect is the relevant mechanism for spin generation even in 1D nanowires.

\begin{flushleft}
{\bf Topological mesoscopic physics:}
\end{flushleft}
\vspace{-2mm}

Yet another interesting direction in the TI device research is the mesoscopic physics in TIs. In this regard, the size quantization in TI nanowires leads to a gap opening and the formation of spin-degenerate subbands\cite{Zhang2010,Bardarson2010,Cook2011}. The gap at the Dirac point as well as the spin degeneracy of the subbands can be manipulated by a magnetic flux $\Phi$ threading along the nanowire. The 1D bands formed in a TI nanowire are indexed by the angular-momentum quantum number $\ell$. In the presence of $\Phi$, the energy dispersion is given by\cite{Zhang2010}
\begin{equation}
 E_\ell(k)=\pm\hbar v_\mathrm{F} \sqrt{k^2+\left(\frac{\ell-(\Phi/\Phi_0)}{R}\right)^2},
 \label{eqn:Elk}
\end{equation}  
where $v_\mathrm{F}$ is the Fermi velocity, $R$ is the radius, and $\Phi_0=hc/e$ is the flux quantum. The Berry phase coming from the spin-momentum locking makes $\ell$ to take half-integer values $\pm \frac{1}{2},\pm \frac{3}{2},\dots$, making the 1D band to be gapped in zero field. 
The realization of this peculiar subband structure in TI nanowires was recently confirmed in an experiment observing gate-voltage-dependent resistance oscillations\cite{Muenning2021}.

In Eq. (1), the flux of $\Phi = \frac{1}{2}\Phi_0$ cancels $\ell = \frac{1}{2}$ and gives rise to a non-degenerate gapless band --- this band is spinless, which makes it eligible for hosting MZMs when superconductivity is induced.
The dependence of this subband structure on $\Phi$ leads to Aharonov-Bohm (AB)-like oscillations in the nanowire resistance\cite{Zhang2010,Bardarson2010,Peng2010}, when the magnetic field is applied along the nanowire. For example, in an ideal case of the Fermi level $E_{\rm F}$ located at the Dirac point, the density of states (DOS) at $E_{\rm F}$ is zero when $\Phi = n \Phi_0$ with $n$ = 0, $\pm$1, ..., while the DOS becomes finite at $E_{\rm F}$ when the gap is closed at $\Phi = (n + \frac{1}{2})\Phi_0$. This periodic change in the DOS at $E_{\rm F}$ is the mechanism of the AB-like oscillations in TI nanowires. Therefore, when $E_{\rm F}$ is gate-tuned to the bottom of one of the subbands, the AB-like oscillations should show a $\pi$ phase shift. This phase shift as a function of gate voltage has been reported\cite{Cho2015,Jauregui2016}, although the observed phase shift was quite irregular. 

The tunability of the 1D band structure including its spin degeneracy would lead to more exotic mesoscopic effects. For example, application of an electric field by gating, which breaks inversion symmetry, has been theoretically shown to lift the degeneracy in zero magnetic field. Further lifting the Kramers degeneracy with a moderate magnetic field produces a band structure which is particularly suitable for hosting MZMs in the presence of unavoidable disorder\cite{{Legg2021}}, as already mentioned above. 
Even without the Majorana context, the interplay between the Berry phase and various types of symmetry breaking in TI nanostructures offers an interesting playground for mesoscopic physics and quantum transport including nonreciprocal response\cite{Legg2021b}.

\section{Fabrication of TI devices}

In the fabrication of TI devices, there are three main challenges --- realizing and preserving the insulating bulk, protecting and accessing the surface states at the same time, and being able to control and manipulate the surface states. 

\begin{flushleft}
{\bf Realizing and preserving the insulating bulk:}
\end{flushleft}
\vspace{-2mm}

Devices based on TIs are most exciting due to the peculiar properties of their conducting surface states. Their contribution can be enhanced not only by using thin films and exfoliated flakes to achieve a large surface-to-bulk ratio, but also by tuning the materials growth such that the chemical potential lies within the bulk band gap. 
Due to naturally-occurring self-doping, most of the TI compounds such as the binary tetradymite Bi$_2$Se$_3$, Bi$_2$Te$_3$, or Sb$_2$Te$_3$ conduct in their bulk. Nevertheless, the experimental challenge to realize a truly insulating bulk has been overcome by ``compensation'' of dopants, which can be achieved, for example, in the solid-solution of tetradymite compounds having opposite types of naturally-occurring carriers\cite{Ando2013}. Two widely-used  materials belong to this class --- (Bi$_x$Sb$_{1-x}$)$_2$Te$_3$ (hereafter called BST)\cite{JZhang2011} for thin films and Bi$_{2-x}$Sb$_{x}$Te$_{3-y}$Se$_y$ (called BSTS) for bulk crystals\cite{Ren2011}.

Bulk-insulating BST thin films can be grown with molecular beam epitaxy (MBE) by tuning of flux ratios and growth temperatures\cite{JZhang2011}. Because the growth proceeds in the van-der-Waals epitaxy mode\cite{Ando2013}, the lattice-constant matching of the substrate is not required for the epitaxy. While silicon wafers are compatible with well-established industry tools and allow for epitaxial growth of BST\cite{Schueffelgen2019}, sapphire (Al$_2$O$_3$) wafers tend to give better film qualities\cite{Yang2014}. 

Due to the layered structure and weak van-der-Waals bonds of 3D TIs of the tetradymite family, exfoliation of melt-grown bulk crystals of BSTS into thin flakes is another route to obtaining high-quality bulk-insulating TIs ready for device fabrication (see Fig.~\ref{JJ}a as an example). A major advantage of flakes over thin films is the flexibility in the choice of substrates to exfoliate on. Conducting substrates coated with an insulating dielectric, e.g. doped Si coated with SiO$_2$, can be used for simple realization of a gate. For device fabrications based on exfoliated flakes, it is helpful to use prefabricated substrates patterned with an array of markers for coordination in lithography (see Fig.~\ref{fab}b). The thickness of exfoliated flakes can be judged by optical microscopy based on the shade of the flake in the bright field and the contrast of its edge in the dark field (see Fig.~\ref{fab}b) in the crucial thickness range of 5--30 nm.

\begin{figure*}[t]
 \centering
 \includegraphics[width=0.9\linewidth]{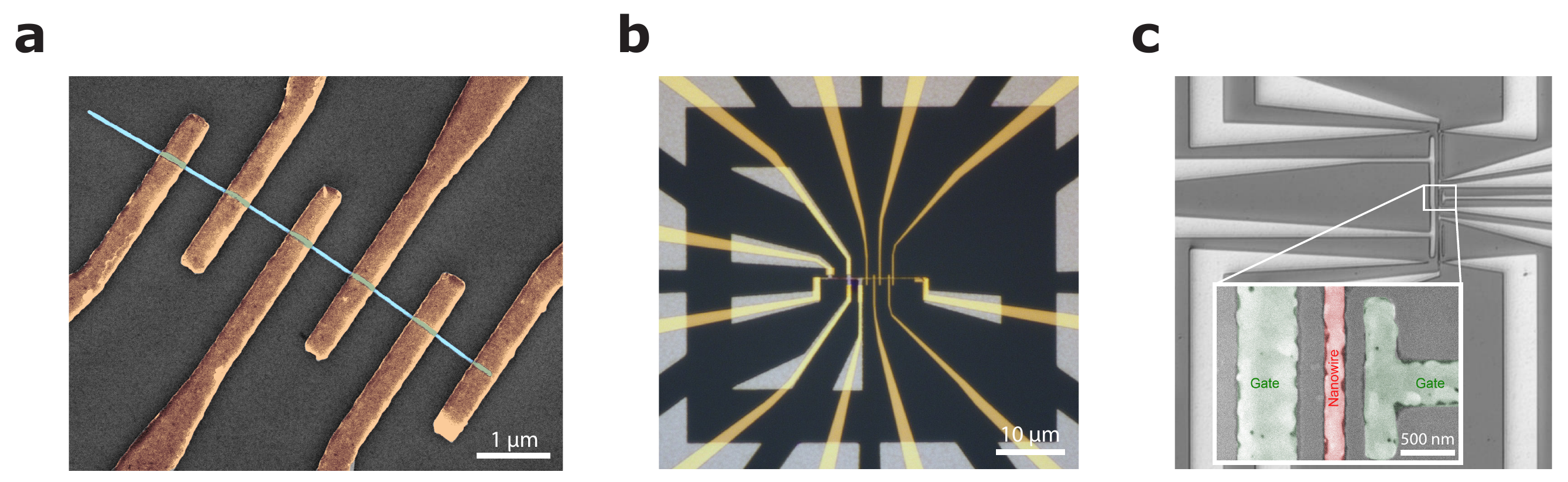}
 \caption{\textbf{Topological insulator nanowire devices.} \ 
 \textbf{a}, False-color scanning electron micrograph of a device based on VLS-grown nanowire (colored light blue) for normal-state transport measurements using electrodes (brown) made from 5\,nm Pt + 40\,nm Au.  
 \textbf{b}, BST thin film (gray) etched into the shape of a nanowire. Contacts are either realized based on etched leads (5 left contacts \& right-most contact) or by subsequent fabrication of Pt/Au finger contacts. \textbf{c}, Nanowire device made by selective area growth. A BST film grows within the prepatterned trenches only (white areas of the main image). Multiple leads, but also gates (green areas in the false-color scanning electron micrograph shown in the inset), can be realized from BST in the same run.  
 }
 \label{NW}
\end{figure*}

Since a major benefit of devices is to have the possibility of gate-tuning the chemical potential, the large density of states coming from conducting bulk would prevent any efficient tuning due to screening, spoiling this benefit. Therefore, it is crucial to retain the insulation of the bulk upon fabrication by not heating the sample too much during the resist baking or depositions of metals and dielectrics, so that the chemical composition of the TI is kept intact.
For example, in the cases of BST films and BSTS flakes, the maximum process temperature is 120$^{\circ}$C.

\begin{flushleft}
{\bf Protecting and accessing the surface states:}
\end{flushleft}
\vspace{-2mm}

It is important to note that the surface of BST and BSTS are oxidized in ambient atmosphere. The oxidation dynamics depends on the chemical composition of the TI material and its surface morphology\cite{Volykhov2016,Volykhov2018}. For strongly-terraced MBE thin films, surface oxidation is more relevant than for exfoliated flakes that are almost atomically-flat. This oxidation causes electron doping, but besides that, there has been no evidence that oxidation degrades the electron mobility in the surface states\cite{Taskin2011} --- most likely, the surface states just migrate beneath the oxide layer without experiencing enhanced scattering. In this sense, the oxide layer protects the surface states.

On the other hand, the oxide layer poses a problem upon taking ohmic contacts or inducing superconductivity through a metal/superconductor deposited on the surface. To remove the oxide layer, wet etching (described later) is preferred over dry etching. Note that after the etching, sizable oxidation typically occurs within minutes to hours at ambient conditions, so the air exposure of the device under fabrication needs to be minimized. An often adopted strategy is to avoid oxidation by capping films {\it in-situ} or right after the growth with minimal air exposure. 
As a capping layer for BST, tellurium layer grown {\it in-situ} in a MBE chamber, or Al$_2$O$_3$ layer grown {\it in-situ} by electron-beam evaporation/sputtering or {\it ex-situ} with atomic-layer deposition (ALD), have proven successful\cite{Hoefer2015, Steinberg2011, Yang2014}. Depositing a thin layer of aluminum and letting it oxidize to form AlO$_x$ is less suitable, as aluminum alloys with TIs\cite{Schueffelgen2017}. 

For making electrical contacts, it is mandatory to remove any kind of capping layer right before metallization. Here, plasma etching, even if performed {\it in-situ}, is not a preferred option, because the etch rate of the tetradymite TIs even at low ion energies is large compared to the capping materials. Thus, even a slight over-etch of the capping layer can quickly remove the whole TI underneath. Larger selectivity is achieved by wet etching, for example, by using aluminum etchant (Type D, Transene) for Al$_2$O$_3$ removal. Remember that oxidation occurs right after removing the capping layer in the crucial interfacial areas if the device is exposed to air. Therefore, the usage of a suitable vacuum transfer box or a gentle {\it in-situ} cleaning procedure performed prior to further fabrication is helpful.

Since BST and BSTS are chemically not very stable, for the fabrication of their devices, established semiconductor-based fabrication processes need to be reconsidered, starting with the choice of process chemicals.
Solvents like acetone and IPA can be used safely and do not degrade the TI's properties. However, TMAH-based developers for photo-lithography considerably etch TI films on a nm/min scale. Thus, when working with very thin films or flakes, the development time must be tuned precisely, or preferably electron beam lithography had better be employed. For thicker films the finite etch rate, in contrast, can be beneficial and reduce the contact resistance via the removal of a native oxide layer during the development.

\begin{flushleft}
{\bf Controlling and manipulating the surface states:}
\end{flushleft}
\vspace{-2mm}

There are mainly three ways to control and manipulate the topological surface states: (i) electrostatic gating, (ii) putting other  materials on the surface, (iii) utilizing size-quantization effects.

\begin{flushleft}
(a) Gate-tuning of the chemical potential
\end{flushleft}
\vspace{-2mm}

For tuning the chemical potential of a TI thin film or a flake by electrostatic gating, the thickness of the film or flake is decisive. Typical resists for lithography impose an upper limit of $\sim$100\,nm. Note that a single gate electrode can, in principle, only tune the electron density of only one of the two (top or bottom) surfaces effectively. Additionally, the interface-dependent band bending effects introduce asymmetry between the two surfaces. Hence, top and bottom surfaces need to be considered as two distinct channels whose chemical potential and thus the conductance differ. To evaluate the contributions of the top and bottom surfaces in the total conductance, a two-band analysis of the magnetic-field dependence in the Hall resistance, $R_{yx}(H)$, from a device such as shown in Fig.~\ref{fab}a, is useful\cite{Ando2013}.

The independence of top and bottom surface channels mentioned above makes it difficult to attain the unique situation where the chemical potential is tuned exactly at the Dirac point, by gating the TI from either the top or the bottom side alone. To achieve full tuning to the Dirac point\cite{Taskin2017,Yang2015,Wang2020}, both sides of the TI need to be gated simultaneously, which is called ``dual gating''. Nevertheless, for thicknesses below about 20\,nm, the two surfaces are close enough such that both of them can be tuned reasonably well by gating from just one side\cite{Yang2014}. Although it is no problem to growth films thinner than $\sim$20 nm by MBE, for flakes, the challenge lies in optimizing the exfoliation technique to obtain thin-enough flakes with acceptable yield and in identifying them by their different optical properties. Note that, even for films and flakes thinner than $\sim$20 nm, the chemical potential may have an offset between top and bottom surfaces, albeit being simultaneously tunable.

Full gate control can only be achieved via dual gating. For thin films, this can be done either by growing the film on a thin SrTiO$_3$ substrate\cite{Chang2013} or by transferring the film from the growth substrate to a SiO$_2$-coated doped-Si substrate via a detaching technique\cite{Yang2015} (see Fig.~\ref{fab}c). The quality of the top-gate dielectric decides the later device performance in terms of hysteresis due to trap states, pinning of the chemical potential by localized states, leakage because of pinholes, and premature breakdown due to defects. In this sense, oxides grown in a dry process are preferable and typically used for the back gate. For TI devices, the growth of the dielectric layer needs to be performed without heating the TI material too much, even though a high-quality gate dielectric often requires a high-temperature growth; in this regard, SiN$_x$ grown by hot-wire CVD\cite{Yang2014}, Al$_2$O$_3$ grown by ALD\cite{Checkelsky2014}, and flakes of h-BN transferred onto TI flakes\cite{Fatemi2014, Wang2020} have proved useful.

\begin{flushleft}
(b) Inducing a superconducting gap
\end{flushleft}
\vspace{-2mm}

The surface states of TIs can be manipulated by interfacing them with a superconductor (SC) or a ferromagnet.
One can open a superconducting gap in the TI surface states by using the superconducting proximity effect, but it requires careful optimization of the interface between the TI and a SC to realize a robust superconducting gap. 
To make the TI surface free from residual oxides or adsorbates upon deposition of a SC layer, successive {\it in-situ} deposition of a SC onto a freshly-grown TI thin film is a possibility\cite{Schueffelgen2017}. However, an epitaxial growth of a SC compatible with post-fabrication of nano-devices has not been achieved. In an {\it ex-situ} process, interface cleaning can be performed by gentle (reactive) ion etching and dilute acid treatment\cite{Ghatak2018}. In addition, {\it in-situ} plasma cleaning at low energies ($\lesssim 90\,$eV) can help to prepare pristine surfaces, but the ion energy and milling time should be tuned carefully to avoid surface degradation. 
In this context, atomic hydrogen cleaning that has been successfully applied to semiconductor/SC interfaces\cite{Gazibegovic2017,Heedt2020} can be an interesting alternative, although there has been no report on its application for TIs at lower substrate temperatures.

Besides physically cleaning the interface, the choice of the proximitizing SC material and/or the adhesion-layer material is crucial for achieving a high interface transparency (see Table 1 for an overview). Not all common SC materials can be used directly on top of TI materials due to potential alloying with the TI. For example, among the popular SC materials, aluminum (Al) alloys with BST(S)\cite{Schueffelgen2017}. Thus, it is often necessary to first coat the TI with an adhesion layer which also acts as a diffusion barrier. Platinum and titanium are commonly used for this purpose. The analysis of the TI/SC interface using a Josephson junction is discussed in the {\bf Box}.

\begin{flushleft}
(c) Inducing a size-quantization gap --- Nanowires
\end{flushleft}
\vspace{-2mm}

As already discussed, when a three-dimensional (3D) TI is made into a nanowire, the size quantization effect causes the Dirac surface states to acquire a gap and split into spin-degenerate subbands. For a nanowire of diameter $R \simeq$ 30 nm, Eq. (1) gives the quantization gap $\Delta_{\rm NW}$ ($\simeq \hbar v_{\rm F}/R$) of the order of 10 meV. 
Suitable nanowires can be obtained in various ways. The vapor-liquid-solid (VLS) growth yields TI nanowires in length of a few $\mu$m of consistent morphology (Fig.~\ref{NW}a)\cite{Peng2010, Muenning2021}. Hydrothermal synthesis\cite{Xiu2011} or exfoliation of bulk crystals\cite{Cho2015} can also result in thin TI nanoribbons.
The approach of etching a continuous thin film into desired nanowires (or a network of them) promises a large design freedom; nanowires with carrier mobilities retained from the pristine film can be obtained by a combination of dry ion etching and wet etching\cite{Legg2021b} (Fig.~\ref{JJ}b and Fig.~\ref{NW}b).
Also, the ``selective area growth'' method (Fig.~\ref{NW}c) of prepatterning a substrate with a growth mask for subsequent epitaxy allows for complex device designs as well and can additionally be extended to employ {\it in-situ} deposition of superconductors by using suspended growth masks\cite{Schueffelgen2019, Rosenbach2020}.

\begin{table*}
\label{tab}
{\small
\begin{tabular}{lcp{5cm}p{6.5cm}}
\toprule
\textbf{material} 	& 	\textbf{adhesion layer\,\,}	&	\textbf{deposition method}	&	\textbf{comment}\\
\midrule 
niobium	(Nb)		&		--			&	UHV sputtering		&	SC, great adhesion, high $T_c$, no epitaxy \cite{Kurter2014} \\
aluminum (Al)		&	Ti, Pt			&	preferred: e-beam evaporation\newline alternative: thermal evaporation	&	SC, good transparency, considerable aging if not capped \cite{Williams2012,Ghatak2018}\\
vanadium (V)		&	Ti				&	e-beam evaporation	&	SC, good transparency, tricky liftoff\\
PdTe$_2$		&	--				& HV sputtering of Pd	&	SC, self-formed upon deposition of Pd \cite{Bai2020}\\
tungsten (W)	& --				& FIB					&	SC, difficult to deposit\cite{Zhang2011, Bhattacharyya2018}\\
lead (Pb)			&	--				& thermal evaporation	&	SC, unpopular material for microfabrication\cite{Yang2012, Qu2012}, topological proximity effect \cite{Trang2020}\\
EuS & -- & MBE & FM insulator, epitaxial growth on Bi$_2$Se$_3$ \cite{Katmis2016}\\
Zr$_{1-x}$Cr$_x$Te & -- & MBE & FM insulator, epitaxial growth on BST to achive QAH effect \cite{Watanabe2019}\\
Permalloy (Fe$_{20}$Ni$_{80}$) & -- & thermal evaporation & FM metal, used for spin detection \cite{Yang2016}\\
cobalt (Co) & -- &  thermal/e-beam evaporation & FM metal, used for spin detection \cite{Li2014}\\
iron (Fe) & -- &  e-beam evaporation / MBE & FM metal, used for spin detection \cite{Li2014}\\
MgO & -- & e-beam evaporation / sputtering & Insulator, tunnel barrier for spin detection \cite{Li2014}\\
Al$_2$O$_3$ & -- & ALD is preferred & Insulator, used for capping and tunnel barrier, chemically removable \cite{Li2014}\\
tellurium (Te) & -- & MBE & Insulator, capping BST film in the MBE chamber, removable by heating \cite{Kayyalha2020}\\
selenium (Se) & -- & MBE & Insulator, capping Bi$_2$Se$_3$ film in the MBE chamberr, removable by heating\\

\bottomrule
\end{tabular}
}
\caption{\textbf{Materials used for interfacing with a topological insulator.} In TI devices, superconductors (SCs) and insulating ferromagnets (FMs) are used for inducing a superconducting gap and a magnetic exchange gap, respectively, in the surface states. Metallic FMs are used for spin detection. The capping layer to protect the surface states should be easily removable. }
\end{table*}

\begin{flushleft}
(d) Interfacing with ferromagnets
\end{flushleft}
\vspace{-2mm}

Ferromagnets in contact with the TI surface offer two types of functionalities. One is to break time reversal symmetry in the TI surface states through ``magnetic proximity effect'' and to open up a magnetic exchange gap at the Dirac point. The other is spintronic functionalities, either to detect the spin polarization on the TI surface or even to present magnetization switching due the spin-orbit torque exerted from the TI\cite{Hellman2017}. Insulating ferromagnets are used for the magnetic proximity effect, while metallic ferromagnets are used for spintronic devices.

For the magnetic proximity effect, the ferromagnetic insulator EuS grown on Bi$_2$Se$_3$ has been used for devices to detect chiral currents at magnetic domain boundaries\cite{Wang2015} or to induce ferromagnetism in the TI surface at room temperature\cite{Katmis2016}. Also, magnetic proximity effect is a cleaner way to induce ferromagnetism in the TI surface states than magnetic doping to maintain a high carrier mobility, and the QAH effect has recently been achieved by interfacing the ferromagnetic insulator Zn$_{1-x}$Cr$_x$Te to BST\cite{Watanabe2019}. For efficient magnetic proximity effects, using the MBE technique to realize epitaxial interface is crucial\cite{Katmis2016,Watanabe2019}.

As already mentioned, in spintronic devices based on TIs, ferromagnets are used as a spin-voltage detector to probe the current-induced spin polarization in the spin-momentum locked surface states\cite{Li2014,Yang2016}, and for this purpose, the ferromagnetic metals must be employed, preferably a half-metal. Thermally-evaporated Permalloy is commonly used for this purpose. For the spin-voltage detection, non-Ohmic, high resistance contact is preferred to compensate for the impedance mismatch at the interface, and a tunnel barrier is typically inserted in between the TI and the ferromagnet to enhance the spin voltage. For TIs, Al$_2$O$_3$ or MgO is often used as the tunnel barrier material.

The current-induced spin polarization can be used for switching the magnetization of a ferromagnetic metal in direct contact with the TI surface. Such devices use spin-orbit torque, which requires a good metallic contact. The first demonstration of the switching operation was performed in a epitaxially-grown bilayer of BST and ferromagnetic Cr-doped BST\cite{Fan2014}. A very efficient spin-orbit torque switching has been reported\cite{Khang2018} for a device based on Bi$_{0.9}$Sb$_{0.1}$ and MnGa.

\section{Conclusion}

TI devices offer a lot of opportunities for fundamental discoveries, in particular in the areas of Majorana physics, topological magnetoelectric effects, spintronics, and topological mesoscopic physics. Currently, the factors that hold us back are mainly related to materials: the quality of the interface to a superconductor or a ferromagnet had better be improved, preferably by finding a right combinations to allow for epitaxial interface and a high electronic transparency; furthermore, the properties of the TI material itself had better be improved for fewer Coulomb disorder and a higher surface carrier mobility. Since the epitaxial interface preparation usually requires {\it in-situ} successive growths of the upper and bottom layers, developing a suitable nanofabrication technology for devices involving such an epitaxial interface is also important --- this requires either a post-growth etching process or an {\it in-situ} nanostructuring process to be combined with the growth\cite{Schueffelgen2019}.

An important practical issue is the reproducibility. In the case of TI devices, there are three main sources that affect the reproducibility of the results: interface cleanliness, level of bulk insulation, and distribution of Coulomb disorder. The first two issues can be improved with a tighter control of the fabrication process, but the last one is uncontrollable and unavoidable in strongly compensated TIs. In this regard, it would be desirable to find a new TI material which is bulk-insulating without compensation.

\begin{flushleft}
{\bf Acknowledgements:}
\end{flushleft}
\vspace{-3mm}

We thank J. Feng, M. R\"ossler, D. Fan, L. Dang, F. M{\"u}nning, G. Lippertz and A. Taskin for providing device pictures, and Y. Tokura, M. Kawasaki, H. F. Legg, D. Gr\"utzmacher, P. Sch\"uffelgen, and F. Yang for useful discussions. This work has received funding from the European Research Council (ERC) under the European Union's Horizon 2020 research and innovation programme (grant agreement No 741121) and was also funded by the Deutsche Forschungsgemeinschaft (DFG, German Research Foundation) under CRC 1238 - 277146847 (Subprojects A04 and B01) and AN 1004/4-1 - 398945897, as well as under Germany's Excellence Strategy - Cluster of Excellence Matter and Light for Quantum Computing (ML4Q) EXC 2004/1 - 390534769.

\renewcommand{\thefigure}{B\arabic{figure}} 
\renewcommand{\thetable}{B\arabic{table}} 

\setcounter{figure}{0}
\setcounter{section}{0}

\vspace{2mm}
\begin{flushleft} 
{\bf Box: Analysis of TI/superconductor devices}
\end{flushleft} 
\vspace{-2mm}

The success of interfacing a TI with a superconductor needs to be judged from the device characteristics, mostly by performing transport experiments in a dilution fridge. Useful information can be gained from Josephson junctions, where two superconducting electrodes are placed on top of the TI with a small gap in between them, such that a SNS junction is formed through the TI in a planar geometry (Fig.~\ref{JJ}). When the TI material is bulk-insulating, the gap between the electrodes should be narrower than $\sim$80 nm. 
The width of the superconducting leads should be kept small enough (shorter than the London penetration depth) such that Abrikosov vortices will not penetrate. Abrikosov vortices formed near the junction would make it difficult to observe the Fraunhofer pattern in the magnetic-field dependence of the critical current $I_c$, which gives insight into the homogeneity of the SC/TI interface and serve as a proof of proximity-induced superconductivity\cite{Ghatak2018}. 

\begin{figure}[h]
 \centering
 \includegraphics[width=0.8\linewidth]{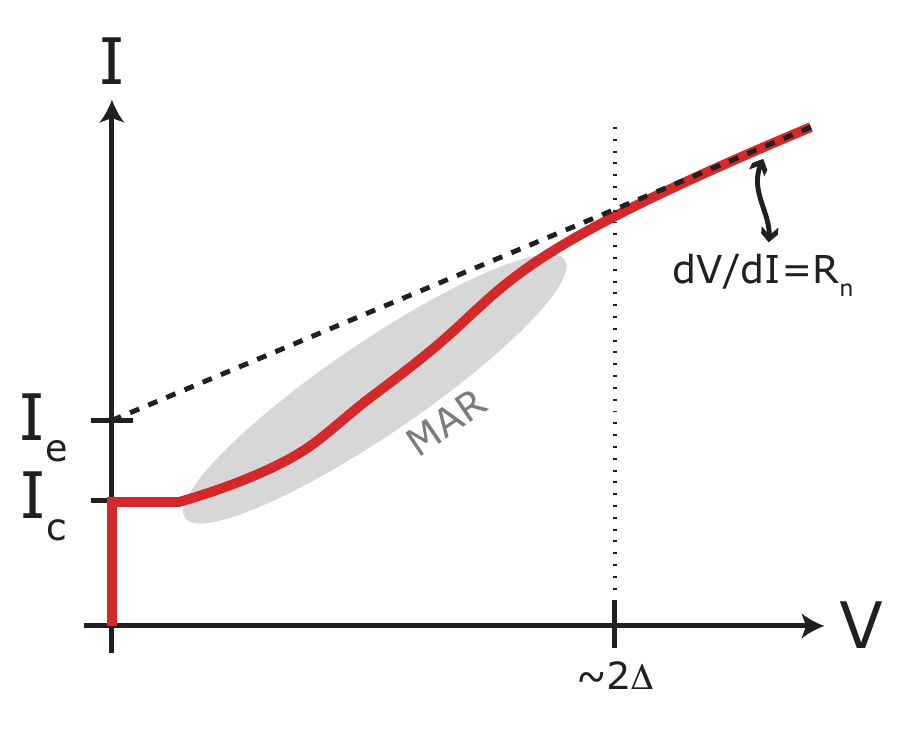}
 \caption{\textbf{Schematics of the $I$-$V$ curve of a SC/TI/SC Josephson junction.}\ For currents larger than the critical current $I_\mathrm{c}$, the measured $I$-$V$ curve (red line) may contain signatures of multiple Andreev reflections (MAR) up to a voltage of $2\Delta$, where $\Delta$ is the induced superconducting gap. For voltages $V>2\Delta$ a linear regime is entered, characterized by the normal state resistance $R_n$. Extrapolating this linear dependence yields the excess current $I_\mathrm{e}$ as the current-axis intercept.}
 \label{IV}
\end{figure}

In addition to the Fraunhofer-pattern, the {\it I-V} characteristics (Fig.~\ref{IV}) of the junction should be measured to perform the following analyses:\newline
(i) Comparison of the $I_c R_n$ product to the superconductor's bulk gap $\Delta_\mathrm{SC}=1.76\, k_\mathrm{B}\, T_\mathrm{c}$ \newline
(ii) Estimation of the induced superconducting gap $\Delta_\mathrm{ind}$ by indexing the multiple Andreev reflections observed at $V_n=2\Delta_\mathrm{ind}/e n$ in the resistive state\cite{Octavio1983, Xiang2006}, and compare it with $\Delta_\mathrm{SC}$ \newline
(iii) Estimation of the transparency\cite{Blonder1982} between the proximitized TI (underneath the superconducting electrodes) and the pristine TI part (in the junction gap)  from $\Delta_\mathrm{ind}$, excess current $I_{\rm e}$, and the normal-state resistance $R_{\rm n}$ using the theory of Octavio et al.\cite{Octavio1983} and from the temperature dependence of the critical current\cite{Galaktionov2002, Schueffelgen2019}

The TI-based Josephson junctions having a high interface transparency judged from the above analysis tend to present signature of 4$\pi$-periodic current-phase relationship in their ac Josephson effects\cite{Wiedenmann2016, Bocquillon2017, Li2018, Schueffelgen2019, Deacon2017} discussed in the main text. This is the first indication of a possible realization of topological superconductivity. Ultimately, the existence of non-Abelian MZMs should be documented by making a Majorana qubit device to perform braiding and to read out the resulting change in the qubit state.

\clearpage


\end{document}